# A common framework for aspect mining based on crosscutting concern sorts

Marius Marin, Leon Moonen and Arie van Deursen

# A common framework for aspect mining based on crosscutting concern sorts

**Marius Marin**
Delft University of Technology
The Netherlands
A.M.Marin@tudelft.nl

**Leon Moonen**
Delft Univ. of Technology & CWI
The Netherlands
Leon.Moonen@computer.org

**Arie van Deursen**
Delft Univ. of Technology & CWI
The Netherlands
A.vanDeursen@tudelft.nl

**Abstract**

*The increasing number of aspect mining techniques proposed in literature calls for a methodological way of comparing and combining them in order to assess, and improve on, their quality. This paper addresses this situation by proposing a common framework based on crosscutting concern sorts which allows for consistent assessment, comparison and combination of aspect mining techniques. The framework identifies a set of requirements that ensure homogeneity in formulating the mining goals, presenting the results and assessing their quality.*

*We demonstrate feasibility of the approach by retrofitting an existing aspect mining technique to the framework, and by using it to design and implement two new mining techniques. We apply the three techniques to a known aspect mining benchmark and show how they can be consistently assessed and combined to increase the quality of the results. The techniques and combinations are implemented in FINT, our publicly available free aspect mining tool.*

## 1. Introduction

Aspect mining research aims at providing techniques and tools that support the identification of crosscutting concerns in existing code. Such concerns are of interest since they are particularly difficult to manage and understand due to their specific lack of modularization and locality.

With a growing number and variety of mining techniques proposed in literature, it becomes increasingly important to aim at consistency and compatibility between these techniques and their results. Such properties would allow for a systematic evaluation of the techniques, an assessment of results and the combination of techniques to improve quality.

However, most mining techniques rely on heterogeneous descriptions of the crosscutting concerns they aim to identify and the steps to be taken to map their results onto potentially associated concerns. In some cases, the description of the discovered concerns is specific to the context into which they were encountered, and explained through other, better known, examples of crosscutting functionality (e.g., CORBA Portable Interceptors[1] are described as "observer style entities" [14]). Quite often, the mining techniques focus on generic symptoms of crosscuttingness, like tangling or scattering, instead of exploiting specific characteristics of the particular types of concerns they aim to identify. In addition, there is little consistency in describing results and concerns, which makes it hard to compare or combine the results.

Previous experiments aimed at comparing and combining aspect mining techniques [2] show that a significant challenge rises from the lack of a sound definition of crosscutting concerns. This leads to the following (hypothetical but likely) evaluation scenario: One technique describes its results through the participants in an implementation of the Observer pattern that are crosscut by the super-imposed roles of Subject and Observer [3]. A second technique reports results related to the same instance of the pattern, but identified through the elements implementing the crosscutting mechanism of the observers-notification (that is, the methods changing the state of the Subject object consistently invoke a notification method). By interpretation, the human analyzers agree on ad-hoc convergence rules of the results: the Observer pattern instance is counted as a common finding as results from both techniques are valid, and the Observer is a well known example of crosscuttingness. Each technique can further explain how the implementation of the Observer is related to its own identification mechanism.

The problems with the sketched scenario are apparent: the convergence relies on an inconsistent level of granularity for the reported findings, as the Observer implementation comprises distinct crosscutting concerns that the techniques identify. The results require a tedious manual correlation effort as they do not (always) overlap directly but are related by the design decisions they implement. Moreover, the approach requires that, despite their inconsistency, detailed descriptions of results and associated concerns are present. In practice, however, such descriptions are often not available.

To address these issues, we identify a set of requirements for systematic aspect mining aimed at ensuring consistency and compatibility in identification of crosscutting concerns and description of the mining results. These requirements form the basis of a common framework for aspect mining. They comprise a clearly defined *search-goal* for the mining technique, descriptions of the rules for mapping the mining results onto the description of the concerns targeted by the technique, and objective metrics for assessment.

---
[1] Object Management Group - CORBA v3.0.3 specification

Contributions of this paper can be summarized as follows:

- We present a common framework that defines a systematic approach to aspect mining (Section 2);
- We introduce two new aspect mining techniques and show how these and a previously proposed technique conform to the proposed framework (Section 3);
- We apply the three techniques to a common benchmark, both individually and in combination, and assess the results and approach (Sections 6 and 7);
- We provide a high level of automation for the techniques and their combination through tool support (Section 5).

The final sections of the paper present related work, draw some conclusions and discuss future work.

## 2. Common aspect mining framework

The focus of this work is on systematic aspect mining for *generative* techniques: approaches that identify program elements which participate in a crosscutting concern based on source code characteristics, without using domain knowledge about the system that is analyzed. The identified elements are known as crosscutting concern *seeds*.

Most generative aspect mining techniques contain an automatic step in the analysis. The results of this step are *candidate-seeds* (or *candidates*): results which are proposed as seeds by the tool, but still require human inspection and validation. Rejected candidates are *non-seeds* (false positives).

To ensure consistent and systematic aspect mining, we identify a number of requirements for aspect mining techniques. One of these requirements is to define the targeted kinds of crosscutting concerns; that is, the *search-goal* of the technique. We propose to define search-goals using our earlier defined classification of crosscutting concerns in sorts [9, 10].

### 2.1. Crosscutting concern sorts

An important limitation of aspect mining comes from lacking a definition of crosscutting concerns. Such a definition would allow to clearly specify the search-goals of a technique and the mapping between these goals and the actual results. Without this definition, aspect mining techniques have to resort to ad-hoc descriptions of their goals and output and sometimes even omit a detailed specification of their findings and the associated crosscuttingness.

A first step towards overcoming this limitation is a consistent system for addressing and describing crosscutting concerns. To this end, we propose the use of *crosscutting concerns sorts*, a classification system for crosscutting functionality presented in our earlier work [9, 10].

Crosscutting concern sorts are *atomic* descriptions of crosscutting functionality. They are characterized by a number of properties common to all the instances of the sort: (1) a generic description of the sort (i.e., the sort's *intent*), (2) a specific implementation idiom of the sort's instances in a non-aspect-oriented language (i.e., sort's specific *symptom*), and (3) an atomic aspect language *mechanism* to modularize concrete instances of the sort.

Table 1 shows a selection of four crosscutting concern sorts. They are described by their defining properties and by a number of concrete instances. For example, the roles superimposed to participants in a typical implementation of the Observer pattern (the concrete Subject and the Observer roles) are instances of the Role superimposition sort. Similarly, the mechanism of consistently notifying Observer objects of changes in the Subject's state by invoking a notification-method, is an instance of the Consistent behavior sort.

The classification of crosscutting concerns based on sorts ensures a number of important properties for consistent aspect mining: first, the atomicity of the sorts ensures a consistent granularity level for the mining results; second, sorts describe the relation between concrete instances and the associated crosscutting functionality; third, sorts provide a common language for referring to typical crosscuttingness, and hence for defining the search-goals of an aspect mining technique.

### 2.2. Defining the common framework

We propose a common framework for aspect mining that defines a systematic approach to identify crosscutting concerns. The framework is aimed at ensuring consistency of the mining process and compatibility of results. This compatibility would further allow for assessment and combination of mining techniques and results. Conformance with the framework requires a developer of aspect mining techniques to:

**Define the search-goal of the mining technique.** An aspect mining technique has to define its search-goal in terms of kinds of crosscutting concerns that the technique aims to identify. We use the classification system based on crosscutting concern sorts to define search-goals.

**Describe the representation of the mining results.** An aspect mining technique has to define and describe the format for presenting the results of the automatic mining process (i.e., the candidate-seeds). Common formats would typically resemble the specific implementation of the crosscutting concerns targeted by the mining technique (i.e., the sort's implementation idiom).

**Define mapping between mining results and goals.** The mining technique has to define how the candidate-seeds map onto the targeted crosscutting concerns (i.e., the implementation idiom of the targeted sort). This mapping forms the relation between mining results and potentially associated concerns. Furthermore, it describes how we should understand and reason about the candidate-seeds, and how we can expand them into complete crosscutting concern implementations.

Candidate-seeds that cannot be mapped are rejected.

**Define metrics to assess mining technique and results.** We distinguish three metrics: (1) *precision*, (2) *absolute recall*, and (3) *seed-quality* metric [7]. The first metric evaluates the quality of the whole set of candidate-seeds generated by the mining technique. The second counts the absolute number of identified concern seeds. The last metric aims at describing each candidate-seed and providing a measure of the effort required for reasoning about the candidate. Candidates with a low value of the quality metric will typically be dismissed.

| Sort | Intent | Object-oriented Idiom | Aspect mechanism | Instances |
|---|---|---|---|---|
| Consistent Behavior | Implement consistent behavior as a controlled step in the execution of a number of methods that can be captured by a natural pointcut. | Method calls to the desired functionality | Pointcut and advice | Log exception throwing events in a system; Wrap/Translate business service exceptions [8]; Notify and register listeners; Authorization; |
| Contract enforcement | Comply to design by contract rules, e.g., pre- and post-conditions checking | Method calls to method implementing the condition checking | Pointcut and advice | Contract enforcements specific to design by contract |
| Redirection Layer | Define an interfacing layer to an object (add functionality or change the context) and forward the calls to the object | Declare a routing layer (decorator/adapter), and have methods in this layer to forward the calls | Pointcut and around advice | Decorator pattern, Adapter pattern [5]; Local calls redirection to remote instances (RMI) [12]; |
| Role superimposition | Implement a specific secondary role or responsibility | Interface implementation, or direct implementation of methods that could be abstracted into an interface definition | Introduction mechanisms | Roles specific to design patterns: Observer, Command, Visitor, etc.; Persistence [8] |

**Table 1. Sorts of crosscuttingness.**

These three metrics form the core set that is used for assessment; however, this set can be extended with other metrics provided that they are generally applicable.

Optionally (but recommended) the mining technique should provide *guidelines for improving the metric values*. Such improvements can be made, for instance, through combinations of techniques. For example, precision can be improved by combining techniques with the same search-goal; absolute recall can be improved by combining techniques with different goals. Seed-quality would typically be improved by generating results that better overlap with the implementation of their associated crosscutting concerns, and hence have a higher confidence level.

## 3. Three aspect mining techniques

In this section, we describe three techniques for identifying crosscutting concern seeds and how they conform with our framework for systematic and consistent aspect mining. One of these techniques, Fan-in analysis, is a previous contribution, while the other two techniques are contributions of this work. This shows how existing techniques can be retrofitted to the framework and how new techniques can be designed based on the framework's structure. Experiments that apply these techniques to a common case are discussed in Section 6.

### 3.1. Fan-in analysis

Fan-in analysis is a mining technique aimed at identifying crosscutting concerns whose implementation consists of a large number of scattered invocations of specific functionality implemented by a method. The number of distinct call sites gives the *fan-in* metric of the method invoked. The analysis reports methods with large values of their fan-in metric as candidate-seeds. The seeds found using Fan-in analysis typically correspond to crosscutting concerns refactorable by an aspect-oriented *pointcut and advice* mechanism that exists, for example, in AspectJ: the aspect solution captures the calls sites in a pointcut definition and triggers the automatic execution of the method with a high fan-in value at these call sites.

Fan-in analysis can identify a number of crosscutting concern sorts. For example, one type of concerns are mechanisms for consistent notification, such as Observer pattern implementations, consistent logging or tracing operations, exception handling and wrapping, etc. Another type of concerns that are identified by Fan-in analysis is role superimposition.

To improve assessability, we will differentiate between various Fan-in analyses based on the concern sort(s) that are actually targeted by a particular analysis. This will allow us to distinguish between intended and unintended discoveries.

In this paper we will focus on Fan-in analysis aimed at identifying Consistent behavior or Contract enforcement. When we describe properties particular to this analysis, we will refer to it as Fan-in$_{CC}$. In terms of our common framework, it can be defined as follows:

**Search goal** Instances of the Consistent behavior or Contract enforcement sorts.

**Presentation** Results are call relations, described by a callee and a set of callers.

**Mapping** The method with a high fan-in value (the callee) maps onto the method implementing the crosscutting functionality, and the callers of the method correspond to the crosscut elements.

**Metrics** We consider three metrics for assessment:

- precision: the percentage of seeds for instances of Consistent behavior or Contract enforcement in the whole set of reported candidates;
- absolute recall: number of identified seeds (i.e., validated candidates);
- seed-quality: the percentage of callers in the reported call-relation that match elements crosscut by consistent invocation of the method with a high fan-in value. Callers that increase the metric value are those that are validated as participants in the implementation of the associated crosscutting concern.

In our previous work we discuss a number of properties that can be analyzed for improving the seed-quality for Fan-in analysis [7]. These include, for instance, structural or call position relations between the callers of a method with a high fan-in value. High quality candidates contain mostly elements that participate in the implementation of a crosscutting concern, and hence are relevant for reasoning about a candidate.

Recall is likely to improve for lower threshold values of the fan-in metric; however, this is also likely to reduce precision.

## 3.2. Grouped calls analysis

Experience shows that crosscutting concern implementations can be closely related, or that a single concern can be implemented by a number of related method calls. Examples include pre- and post-operation notifications, consistent initialization and clean-up of resources, and multi-step set-up operations. Such concerns typically share their intent and crosscut the same elements. We identify them by looking for groups of methods that consistently invoke a same set of callees.

Grouped calls analysis applies formal concept analysis [4, 6] to all calls in the analyzed system in order to find maximal groups of callees that are invoked by the same callers.

**Search goal**  Instances of the Consistent behavior or Contract enforcement sorts.

**Presentation**  The results are concepts, where the grouped callees are the attributes and the callers are the objects in the concept.

**Mapping**  The attributes in the concept (i.e., the callees) map onto methods implementing crosscutting functionality, and the objects in the concept (i.e., the callers) match the crosscut elements.

**Metrics**  We consider the same three metrics for assessment as for Fan-in analysis:

- precision: the percentage of seeds for Consistent behavior or Contract enforcement instances in the whole set of reported candidates;
- absolute recall: number of identified seeds;
- seed-quality: is given by two partial measures: (1) the percentage of callers that are indeed crosscut by a consistent call to a specific functionality and (2) the percentage of callees that are part of the crosscutting concern implementation as assessed by an human analyzer. The value of the metric is obtained by multiplying the partial measures.

Improving the seed-quality for this analysis can target the set of callers for a reported group of callees, similar to Fan-in analysis, as well as the set of grouped callees, by selecting only those callees that are relevant for a potentially associated crosscutting concern.

## 3.3. Redirections finder

Redirections finder is a technique that looks for classes whose methods consistently redirect their callers to dedicated methods in another class. A typical example are implementations of the Decorator pattern [3]: The Decorator class' methods receive calls, optionally add extra functionality, and then redirect the calls to specific methods in the Decorated class.

To detect such a consistent, yet method-specific, redirection concern, the technique looks for classes (C) whose methods (m[i]) invoke specific methods from another class D (D.n[j]). The automatic selection rule is:

*C.m[i] calls D.n[j] and only n[j] from D* and
*D.n[j] is called only by m[i] from C.*

Class C and its redirector methods are reported by the technique if the *number* of methods in C complying with these conditions is above a chosen threshold, and if the *percentage* of methods in C complying with the conditions w.r.t. the total set of methods of C is higher than a second threshold.

**Search goal**  Instances of Redirection layer.

**Presentation**  Set of pair methods from two different classes, related by one-to-one call relations.

**Mapping**  The callers in the reported set match the methods executing the redirection, while their pair callees receive the redirection.

**Metrics**  We consider three metrics for assessment:

- precision: the percentage of Redirection layer seeds in the set of reported candidates;
- absolute recall: number of identified seeds;
- seed-quality: the percentage of redirectors in the reported candidate.

To further improve the quality, we can add a filter that checks for matching names between the callers and callees. This is a common practice for implementing redirectors, although it could also introduce false negatives.

## 4. Combining techniques

Effectiveness of aspect mining can be increased by combining techniques. This section investigates various combinations of techniques and discusses the effect on the combined results.

### 4.1. Improving precision

Precision is measured by the percentage of crosscutting concern seeds in the complete set of candidates reported by the (automatic) mining technique. A straightforward combination of two aspect mining techniques that increases precision is achieved by intersecting their results (i.e., the set of candidates). However, this can be done only when the techniques target the same crosscutting concern sorts, with compatible representations of the results.

Two techniques that satisfy this condition are, for example, Fan-in and Grouped calls analyses. To combine them, we select those results of Fan-in analysis whose callees occur as callees in at least one of the Grouped calls candidates.

### 4.2. Improving absolute recall

To improve absolute recall, we can simply consider the union of the results of different mining techniques. For techniques that target different concern sorts, this union will not contain overlap in the individual results, and the number of seeds for the combination is the sum of the seeds for each technique.

As argued before, another way of improving the absolute recall is by being less selective, by lowering the thresholds. However, this is likely to reduce precision. For Fan-in and Grouped calls analyses, precision can be restored by combining the techniques with the same search-goal, and taking the intersection of their results. The lower thresholds allows for new candidates to be reported and the intersection filters the results so the precision does not drop significantly.

### 4.3. Improving the seed-quality

Combination of techniques targeting same sort's instances also allows for improvement in the seed-quality of the candidates. For example, we can consider the intersection of the results for Fan-in and Grouped calls analyses, selecting the common callees and the common callers of these callees. Thus, for the same value of the threshold for the number of callers, we consider only callees reported by both techniques, and the callers reported by Grouped calls analysis.

Because Grouped calls analysis is the most restrictive of both techniques, the number of callers for a callee is typically lower than for Fan-in analysis. Moreover, the quality of the combined results will be higher than for Grouped calls analysis alone because the combination takes only one callee, and hence we have no false positives grouped together with seeds.

One result of Fan-in analysis can occur in multiple groups of callees reported by Grouped calls analysis: in this case, we select the Grouped calls result for which the callers set has the largest overlap with the set of callers for the Fan-in candidate.

## 5. Tool support

To experiment with the ideas laid out in this paper, we have extended our free aspect mining tool FINT[2] to include automatic support for the three techniques and their various combinations. FINT is available as an Eclipse[3] plug-in. Figure 1 shows part of the functionality provided by the tool, including views for displaying results of each of the three techniques, and the Seeds view where candidates marked as seeds by the user can be managed, documented and made persistent.

Each technique allows for a number of specific, automatic filters, like filters for *utility* elements or *accessor* methods. Utility elements are those that the user considers as irrelevant for analysis. To filter them, the user is presented with the hierarchical structure of the top-level Java element selected for analysis (e.g., a Java project) in which the elements to be ignored can be selected (e.g., all the elements in packages containing JUnit tests). The accessor methods, that is getter and setter methods, are filtered by automatic inspection of either the signature of the methods or their implementation.

The results for each technique are displayed in a dedicated view, following the representation described in Section 3. The views allow for various sorting operations and code inspection from the elements selected by the user in the view. The user can further open and inspect each candidate in a new view, and run a number of analyses for improving the quality of the candidate. These analyses include inspection of various structural relationships between the elements describing a candidate [7].

Support for combining techniques is available, for example, through intersection of the sets of results of two techniques: The views showing the results can be synchronized so common findings are highlighted in the views. For example, the highlighted elements in the Fan-in Analysis View of Figure 1 correspond to methods that are also present in at least one group reported by the Grouped calls analysis. The bold colored elements show candidates marked as seeds by the user. These elements are also shown in the Seeds view.

## 6. Experiment

We apply the aspect mining techniques described above to JHOTDRAW,[4] which has been proposed and used as common benchmark for aspect mining [8, 2]. Detailed results of the experiments discussed in this section are available on-line.[5]

### 6.1. JHOTDRAW

JHOTDRAW is an open-source framework for bi-dimensional drawings editors. The distribution (v 5.4b1) comes with a default drawing application that we also analyze. The system is a show-case for applying design pattern solutions in a Java implementation. Its size is approximatively 20,000 non-comment, non-blank lines of code.[6]

### 6.2. Applied filters

Table 2 shows the filters applied for conducting the mining experiments on the selected case-study. For all techniques, we filter out the (JUnit) test classes delivered with the application; i.e., the methods from the test classes do not occur among the reported candidates of any technique, and methods from these classes do not contribute to the fan-in metric of a method.

---

[2] http://swerl.tudelft.nl/view/AMR/FINT (v0.6)
[3] http://www.eclipse.org/
[4] http://www.jhotdraw.org/
[5] http://swerl.tudelft.nl/view/AMR/CombinationResults
[6] SLOCCount: http://www.dwheeler.com/sloccount/

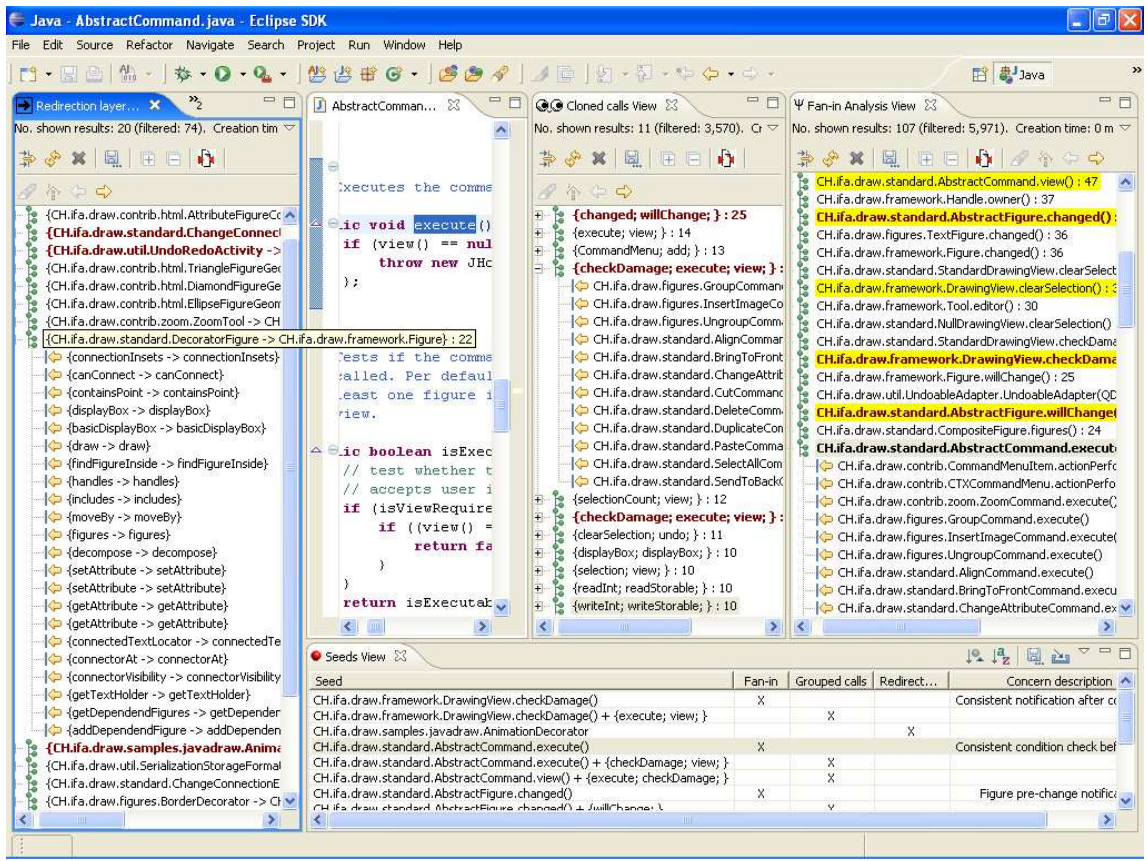

**Figure 1. FINT views**

|  | **Fan-in$_{CC}$** | **Grouped calls** | **Redirection** |
| --- | --- | --- | --- |
| Targeted sorts | Consistent behavior (and Contract enforcement) | Consistent behavior (and Contract enforcement) | Redirection layer |
| Utility filters | Collection wrappers and test classes | Collection wrappers and test classes | Test classes |
| Accessor filters | Accessors by name and implementation | Accessors by name and implementation | - |
| Threshold filters | No. callers: 10 | No. callers: (1)10 and (2)7; No. grouped callees: 2 | No. redirectors: 3; Percentage redirectors: 50% |
| Accepted Seed quality | Above 50% | Above 50% | Above 50% |

**Table 2. Selection conditions applied for the combination experiment.**

Collection wrappers, like *IteratorWrapper* or *SetWrapper*, are also marked as utilities to be filtered from the set of candidates. Similar to the test classes, these wrappers are typically part of dedicated packages (*CH.ifa.draw.util.collections.\**). Collection elements tend to be frequently used in an application, however, in most cases they are not part of a consistent mechanism associated with crosscutting functionality. Filtering these elements is likely to reduce the number of candidates without introducing false negatives.

For Fan-in and Grouped calls analysis, we also filter accessor methods from the set of candidates. The filters check both the signatures of methods (*set\** and *get\** names) and their implementation (i.e., only set a field or return a reference).

A number of threshold values are specific to each case and can be varied by the user to refine the results:

- Fan-in$_{CC}$: the threshold value for the number of callers of a candidate is set to 10, following considerations from previous experiments [8];

- Grouped calls: the first experiment uses a threshold of 10 for the number of callers, this is lowered to 7 for the second experiment; the threshold for the minimal number of callees to be grouped by a candidate is always set to 2;

- Redirections finder: the technique uses two threshold values, the first sets the minimal number of redirector methods in a class to 3, and the second sets the minimal percentage of methods in the class executing the redirection to 50%. Thus candidates reported by this technique will have at least 3 redirector methods, and at least 50% of all their methods execute the required redirection.

```
//CutCommand.execute()
public void execute() {
  // perform check whether view() isn't null.
  super.execute();

  // prepare for undo
  setUndoActivity(createUndoActivity());
  getUndoActivity().setAffectedFigures(
                              view().selection());

  // key logic: cut == copy + delete.
  copyFigures(view().selection(),
                         view().selectionCount());
  deleteFigures(view().selection());

  // refresh view if necessary.
  view().checkDamage();
}
```

**Figure 2. (Simplified) execute method in JHOTDRAW's Command hierarchy.**

Candidates are marked as seeds if they correspond to a crosscutting concern according to the mapping rules of each technique and the seed-quality of the candidate is above 50%.

### 6.3. Individual results

This section shows a number of typical results and metric values for each of the three techniques. Table 3 shows the metrics as well as the values for the combination experiment discussed in the next section.

**Fan-in analysis**  A number of seeds identified by Fan-in$_{CC}$ implement concerns that crosscut the *Command* hierarchy in JHOTDRAW. *Command* classes follow the design described by the pattern with the same name; they implement an execute method that carries out specific activities in response to, for instance, user actions like menu-items selection.

Figure 2 shows the method for executing cut operations in a drawing editor. The method starts with a condition check implemented by the command's super-class (*AbstractCommand*). Similarly, all the (around 20) methods overriding the *AbstractCommand*'s execute method in non-anonymous classes check this condition. The commands then conclude with a notification of the editor's view.

The check- and notification-actions implement two crosscutting concerns scattered over a large number of methods that invoke these actions, and hence increase the value of their fan-in metric. The first invocation is a typical seed for an instance of the Contract enforcement sort, while the latter implements a concern of the Consistent behavior sort.

To calculate the quality of these results, we have to consider all the callers reported by Fan-in$_{CC}$ for each of the invoked actions. Not all the callers, however, belong to the context of the *Command* hierarchy crosscut by the concerns of the two candidate-seeds. For instance, one of the calls to the execute method originates from an action-event han-dler in a *MenuItem* class. The quality of the candidate for the execute method is given by the proportion of the 18 methods crosscut by the reported call, to the whole set of 24 callers. This value is 75%, above the set threshold of 50% for selecting a candidate as a seed.

| Technique | Precision | Absolute recall (no. of seeds) |
|---|---|---|
| Fan-in$_{CC}$ (FI)[7] | 30% (33/109) | 33 |
| Grouped calls (GC$_1$) | 45% (5/11) | 5 |
| Grouped calls (GC$_2$) | 55% (12/22) | 12 |
| Redirection finder (RF) | 92% (12/13) | 12 |
| FI + GC$_1$ | 41% (7/17) | 7 |
| FI + GC$_2$ + RF | - | 51 |

**Table 3. Metric values for individual and combined techniques.**

The Consistent behavior (Contract enforcement) seeds identified through Fan-in$_{CC}$ analysis count 33 methods in the total set of 109 candidates reported. This indicates a precision of around 30% for the targeted sorts, as shown in Table 3.

**Grouped calls analysis**  The two candidates discussed above for Fan-in analysis share the largest part of their callers, and hence are also among the results reported by the Grouped calls analysis. Although the two concerns are distinct, they are related by the set of elements they crosscut (i.e., the *Command* hierarchy). The Grouped calls analysis does not separate the two concerns, but instead allows to put them in a single, shared context.

One of the candidates reported by this technique groups the view and execute methods in the set of callees, together with 14 common callers. Another candidate groups the same two methods, but also the checkDamage method, together with 12 common callers. In the first case, the execute method is the relevant element for the crosscutting concern associated with the reported candidate. The *view* method has no relevance to this concern, and hence it decreases the quality of the candidate. We dismiss the candidate as we select only those whose quality (for the callees group) is above 50%.

On the second case, however, each invocation of the view method occurs together with a call to the checkDamage method, which is a seed for the previously discussed instance of the Consistent behavior sort. In this case, the reported view method is relevant for the crosscutting concern associated with the reported candidate and contributes to the quality metric. For this candidate, the quality metric is 100% as all the grouped callees and callers belong to the implementation of the related crosscutting concern.

In comparison with Fan-in analysis, the number of results and seeds for this technique is lower for the same threshold for the number of callers. This is to be expected, as this technique

---

[7] Results for Fan-in$_{CC}$ in this work are exclusively for the targeted sorts. They differ from results reported in our earlier work [8] because in that case more concern sorts were targeted. All results are documented on the experiment's web-page mentioned earlier (footnote 5).

```
public abstract class DecoratorFigure {
  // ...
  private Figure myDecoratedFigure;

  public TextHolder getTextHolder() {
    return getDecoratedFigure().getTextHolder();
  }

  public Rectangle displayBox() {
    return getDecoratedFigure().displayBox();
  }

  // Forwards draw to its contained figure.
  public void draw(Graphics g) {
    getDecoratedFigure().draw(g);
  }

  // ...
}
```

**Figure 3. (Part of) DecoratorFigure - super-class for Figure objects decorators.**

has more restrictive selection rules for the candidates: a callee should not only have a large number of callers, but it also has to be called together with at least same one other method.

For a lower threshold, the number of seeds is almost doubled, but is still lower than the one reported for Fan-in analysis.

**Redirection layer analysis**  A number of classes in JHOT-DRAW, like *Border-* or *Animation-Decorator*, extend the *DecoratorFigure* class shown in Figure 3, which provides the basic functionality to forward calls to a decorated *Figure* object. This example is a typical instance of the Decorator pattern: Methods in the Decorator classes consistently redirect their callers to dedicated methods of a target object, before or after (optionally) providing additional functionality.

The Redirectors finder candidate for this concern consists of 22 call relations, 3 of which correspond to the methods shown in the figure. Since all reported results implement the redirection concern the seed-quality of the candidate is 100%.

The high precision of this technique has been confirmed by experiments on other case-studies like Tomcat[8] and JBoss[9].

### 6.4. Combinations

**Precision**  An overview of the quality improvements gained by combining Fan-in and Grouped calls analyses is shown in Table 4. Despite a significantly lower absolute recall value (7), we observe that the precision of the combination is higher, as shown in Table 3.

For the case discussed, the combination of techniques proves successful as it provides a better precision than the individual technique (Fan-in$_{CC}$).

---

[8] http://tomcat.apache.org/
[9] http://jboss.org/

**Absolute recall**  Besides taking the union of results, we also run a second experiment by lowering the threshold for the number of callers required for candidates of Grouped calls analysis from 10 to 7 (labeled with $GC_2$ in Table 3). This experiment allows to consider callees that are potentially missed by fan-in analysis due to its higher threshold filter. Most of the results and seeds of the Grouped calls analysis with a lower threshold overlap with the results of Fan-in analysis, however, a number of seeds were found and included in the union set for the three techniques. The size of the union set counts 51 distinct seeds.

**Candidate-seed quality**  The results of the combination of the two techniques targeting the same sort show improved values for the quality metric. Most of the common results have a higher quality for the Grouped calls analysis, and the combination of techniques would retain these high values.

## 7. Discussion

**Retrofitting existing techniques**  A number of existing mining techniques described in literature appear to be suitable for adaption to the proposed framework. The technique for detection of aspectizable interfaces by Tonella and Ceccato [13], for instance, could be described in terms of our common framework as follows:

**Search goal**  Instances of the Role superimposition sort, as described in Table 1.

**Presentation**  Group of methods that belong to or can be abstracted into an interface definition.

**Mapping**  The reported interface and its members map onto elements that crosscut the types implementing them.

**Metrics**  The precision is given by the percentage of correctly identified crosscutting type (interface) definitions. The number of these types gives the absolute recall measure.

The quality metric is given by the percentage of methods in a result that are correctly identified as part of a crosscutting type definition.

Techniques based on clone detection could also prove suitable for identifying instances of Consistent behavior or Contract enforcement. The Grouped calls technique resembles mining based on clone detection but only in some respects: by not considering the position of the calls in the body of the callers, the technique allows to identify related calls that would typically be missed by classical clone detection tools. On the other hand, it can introduce false positives that probably will not be present in standard clone detection. Although a number of aspect mining experiments have been carried out using clone detection [1, 11], there is no report of a complete analysis of all the results produced by clone detection-based techniques, and their total precision and absolute recall.

| Candidate | FI quality | $GC_1$ quality | $FI+GC_1$ quality |
|---|---|---|---|
| framework.DrawingView.checkDamage | 64% | 100% | 100% |
| framework.DrawingView.clearSelection | 55% | 100% | 100% |
| framework.DrawingView.selectionCount | 63% | 83% | 83% |
| standard.AbstractCommand.execute | 71% | 100% | 100% |
| standard.AbstractFigure.changed | 100% | 100% | 100% |
| standard.AbstractFigure.willChange | 100% | 100% | 100% |
| util.UndoableAdapter.undo | 92% | 100% | 100% |

**Table 4. Values of the quality metric for individual and combined techniques.**

**Interpretation of results**  The Consistent behavior and Contract enforcement sorts share their implementation idiom. Thus, instances of both sorts have the same specific symptoms of crosscuttingness. This allows aspect mining techniques to target instances of both sorts together. However, because the differences between the two sorts are only due to their intent, it is difficult to distinguish between them automatically. As a consequence, assessment requires human interpretation.

The Grouped calls analysis builds concepts of callees-callers based on the complete set of methods of the analyzed system. The methods grouped in a concept and to which no filters have been applied give the *extended* result. That is, an extended result also includes accessor or utility methods that we would typically filter before reasoning about a candidate. To reduce the number of false negatives, we can first apply the filters, reason and decide about a candidate, and then investigate all the remaining elements in the extended representation of the results marked as seeds.

Similar extensions can be achieved for the results of the combination of Fan-in analysis with Grouped calls analysis for improving the seed-quality: the combination considers only the callers from the Grouped calls analysis, as this will typically provide a better seed-quality. However, this intersection might ignore relevant elements from the set of callers reported by Fan-in analysis. A common practice in this combination is to reason about the result of the combination for deciding about a candidate, and then extend the analysis to the other callers reported by Fan-in analysis.

**Multiple search-goals**  As mentioned before, the relation between a technique and its search-goal is not exclusive: the same technique can target instances of different sorts if different mappings are defined and applied.

Earlier, our focus for Fan-in analysis was at identifying Consistent behavior or Contract enforcement. However, if we were to employ Fan-in (or Grouped calls) analysis to identify instances of the Role superimposition sort (Fan-in$_{RS}$), we could define the following mapping: the callers of the high fan-in method belong to the implementation of a crosscutting, super-imposed role, and the reported method with a high fan-in value implements functionality dedicated to and accessed from the scattered places implementing the role.

Several instances of Role superimposition are present in JHOTDRAW. A typical example is the persistence concern: The *Figure* elements implement the *Storable* interface that defines (read and write) methods to (re-)store a figure from/to a file. The scattered implementations of these methods for persistence invoke functionality from classes (*StorableInput* and *StorableOutput*) that are specialized in reading/writing specific types of data. The candidates reported by the technique are the methods in the specialized classes together with their callers in the *Storable* hierarchy.

The persistence candidates for Role superimposition instances add to the total number of seeds identified for the analyzed system. However, since these results are not compatible with the Consistent behavior (Contract enforcement) instances discussed earlier, they should be addressed distinctly if the technique is to be compared with another one. This is achieved by explicitly specifying the search-goal, as is done with Fan-in$_{CC}$ and Fan-in$_{RS}$.

The complete results for various crosscutting concern sorts identified using fan-in analysis is discussed and described in previous work [8].

**Tool performance**  Although this work's focus is less on each individual mining technique and more on the common framework to consistently assess and possibly combine mining techniques, we will briefly discuss the performance of the tool. The analysis of the whole JHOTDRAW system for Fan-in analysis requires around 30 seconds on our test machines (Pentium 4 - 2.66 GHz, with 1GB of RAM) running Eclipse 3.1.x under either Linux or Windows OS.

The Grouped calls analysis requires the model built for the Fan-in analysis and takes around 5 minutes to examine all the call relations (approximately 6000 x 6000 elements). This analysis will not scale up very well to systems like Tomcat or JBoss, which have up to 35,000 elements. However, trying to understand such large systems in one iteration is hardly advisable due to the cognitive complexity. We would suggest dividing them into sub-systems comparable in size with JHOTDRAW and gather understanding for each of these systems. Actually, to ensure maintainability, the architecture of the two aforementioned systems is already conveniently split in components that are suitable for analysis in isolation.

The Redirection finder uses the Fan-in model and requires only a few seconds for execution.

**Reproducibility**  To allow for reproducibility of the experiments described in this paper, we provide both the tool and detailed setup elements and results sets on the tool's and experiment's web-pages, indicated in footnotes 2 and 5.

## 8. Related work

Several authors have proposed (and taken) steps towards the comparison and combination of aspect mining techniques [8, 2, 11]. We are not aware of related work on providing a common framework for systematic aspect mining, and consistent combination and assessment of mining techniques.

Shepherd et al. [11] report on machine learning techniques for combining aspect mining analyses. Their approach learns from annotated code and they compare the results of their combination to results of Fan-in analysis [8]. A drawback is the required annotation of crosscutting concerns on some significant system, which is needed for training the tool.

The techniques considered for combination include some that are also used as filtering support in FINT, such as filters for accessor or utility methods. However, the authors of the experiment do not describe their findings in detail nor do they provide rules to consistently associate results of different representations to crosscutting concerns.

In the collaborative AIRCO effort [2], three aspect mining techniques are compared and investigated from the perspective of combination. The techniques include fan-in analysis, dynamic analysis of execution traces, and analysis of shared identifiers in signatures of program elements. Major difficulties in this experiment were caused by heterogeneity in the search-goals of the three techniques and in representation of results. Such experiments require a tedious effort from the participants in the experiment to bring individual results to comparable levels of granularity. Due to such issues, the experiment could focus only on a limited selection of common findings. This motivated the work presented here.

A broader overview of specific aspect mining techniques falls out of the scope of this paper. We refer the interested reader to surveys of related work in our earlier work [8, 2].

## 9. Conclusions

With a growing number of aspect mining techniques and approaches, it is increasingly difficult to consistently assess, compare and combine mining results.

This paper addresses this situation by proposing a common framework to define systematic aspect mining based on crosscutting concern sorts. The framework allows for consistent assessment, comparison and combination of compliant aspect mining techniques. It identifies a set of requirements that ensure homogeneity in formulating the mining search-goals, presenting the results and assessing their quality.

We demonstrate feasibility of the approach by retrofitting an existing aspect mining technique to the framework, and by using it to guide the design and implementation of two new mining techniques. We apply the three techniques to an aspect mining benchmark known from literature and show how they can be consistently assessed and combined to increase the quality of the results. The techniques and combinations are implemented in FINT, our free aspect mining tool which is publicly available.

As future work, we would like to investigate how other existing mining techniques can be retrofitted to the proposed framework and integrated into FINT. We would also like to experiment with other combinations of aspect mining techniques, and analyses for improving the seed-quality, as discussed in [7].

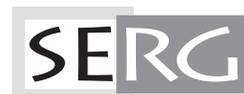